# Control of ultrashort optical electromagnetic pulses in carbon nanotubes at low temperatures

M.B. Belonenko[1], N.G. Lebedev[2], E.N. Nelidina[3], O.J. Tuzalina[3]

[1] *Nanotechnology laboratory, Volgograd Business Institute,*

*Yujno-Ukrainskaya st. 2, 400048, Volgograd, Russia*

[2] *Volgograd State University,*

*University Avenue 100, 400062, Volgograd, Russia*

[3] *Volgograd State Medical University,*

*Pl. Pavshikh Bortsov Square 1, 400131, Volgograd, Russia*

Propagation of the alternating electromagnetic field in a system of zigzag carbon nanotubes in the case of low temperatures and applied external electric fields are considered. The electronic system of the carbon nanotubes is investigated microscopically nonmetering an interaction with a phonon subsystem by the fact that the electromagnetic pulse is critically short. An efficient equation for the vector-potential amplitude of the alternating electromagnetic field is obtained. Solutions of solitons analogs which correspond to solitons in the case of cosine dispersion law for the electronic subsystem have been elicited. The dependences of obtained nonlinear solutions on problem parameters and the applied external electric fields were analyzed. The possibility to control the shape of optical pulse in wide range was shown.

## 1. Introduction

Researches showed that nanotubes had unique properties: very high strength, conduction of semiconductors and metals and other properties providing unlimited possibilities for instance in the microelectronics [1-4]. Because of relational simplicity of the nanotubes structure and their quasi-one-dimensionality these substances are very popular among either the theorists or experimentalists. As expected the nanotubes nonlinear properties of acoustic and electromagnetic nature initiate special interest. For example, the carbon nanotubes nonlinear properties were investigated [5-7] using the reduction of the Korteweg de Vries equation. Moreover, there are points of the carbon nanotubes nonlinear properties in the optical range out of concern. The questions of a CNT nonlinear response to electromagnetic field were explored in works [8, 9]. According to conclusions made from these investigations the nonlinearity is due to both variation of the classical electrons distribution function and the nonparabolic electrons dispersion law. It should be noted that the question of using an approach based on the Boltzmann kinetic equation needs validation whereas an approximation of relaxation time for the ultrashort laser pulses is open to question. There are some questions required further adjustments in spite of solitons existence possibility and dependence of their parameters on the carbon nanotubes parameters were established in works [10, 11]. Note that in case of ultrashort optical pulses it is necessary to use the kinetic equation which differs from researches [10, 11], since the pulse duration is much less than relaxation time scale. It is obvious that the development of a consequent quantum mechanical description based on the microscopic Hamiltonian for the carbon nanotubes electrons system represents an independent academic interest. It becomes particularly current problem in case of low temperatures when electron gas statistic degeneration must be taken into account. Besides, all the aforementioned researches did not neglect an effect of applied alternating electric fields which can exert a significant influence upon the alternating field pulse dynamics. The possibility of an optical pulses shape control by means of external fields is also of a great interest. All the mentioned turned out to the stimulus for this research.

**2. Basic equations**

The electronic structure of carbon nanotubes has been investigated in a large number of woks [12, 13] and as a rule by analyzing the dynamics of π electrons in the framework of the tight-binding approximation. Let us consider alternating electric field, propagating in a system of carbon nanotubes (CNT) in the geometry depicted in Fig.1.

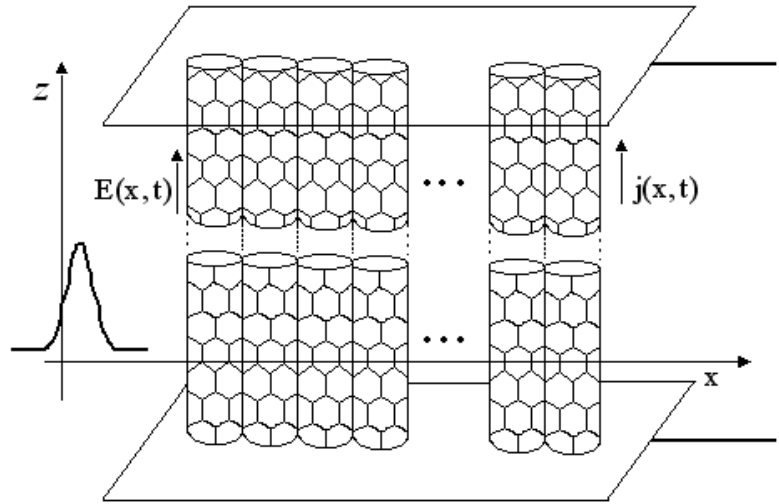

Fig.1. Geometry of the problem. The constant field is parallel to the alternating field.

In case of presence of the external alternating electric field in the gauge $\vec{E} = -\dfrac{1}{c}\dfrac{\partial \vec{A}}{\partial t}$ Hamiltonian of the electrons system has the form:

$$H = \sum_{ps} \varepsilon_s(p - \frac{e}{c}A(t) - \frac{e}{c}A_0(t))a^{+}_{ps}a_{ps} \tag{1}$$

where $a_{ps}^+, a_{ps}$ are electrons creation and annihilation operators with quasimomentum ($p, s$); $A(t)$ is a value of the alternating electromagnetic field vector potential, which has a single component and aligned parallel to the axis of the nanotube; $\varepsilon_s(p)$ is the electrons dispersion law; $A_0(t)$ is the vector potential of the applied external electric field, which is parallel to the alternating field of the electromagnetic pulse. It should be noted that the dispersion law describing the properties of a graphene has the form [14]:

$$E(\vec{p}) = \pm\gamma\sqrt{1 + 4\cos(ap_x)\cos(ap_y/\sqrt{3}) + 4\cos^2(ap_y/\sqrt{3})},$$

where $\gamma \approx 2.7$ eV, $a = 3b/2\hbar$, $b = 0.142$ nm is the distance between the neighboring carbon atoms in the graphene, $\vec{p} = (p_x, p_y)$. The different signs correspond to the conduction and valence bands. In order to derive the dispersion law for a carbon nanotube, it is sufficient to take into account the way of rolling up the graphene sheet into a cylinder and to impose the condition for quantization of the quasimomentum $\vec{p}$ in the direction along the circumference of the carbon nanotube. Thus, for a zigzag CNT which will be investigated for definiteness of the problem, we obtain:

$$\varepsilon_s(p) = \pm\gamma\sqrt{1 + 4\cos(ap)\cos(\pi s/m) + 4\cos^2(\pi s/m)} \qquad (2)$$

where the quasi-momentum $\vec{p}$ is defined as $(p, s)$, $s = 1, 2 \ldots m$ and the nanotube has the type $(m, 0)$.

The Maxwell equations with regard to the dielectric and magnetic properties of carbon nanotubes [15] can be written in the form

$$\frac{\partial^2 \vec{A}}{\partial x^2} - \frac{1}{c^2}\frac{\partial^2 \vec{A}}{\partial t^2} + \frac{4\pi}{c}\vec{j} = 0 \qquad (3)$$

here we disregard the diffraction spreading of the laser beam in the directions perpendicular to the axis of the pulse propagation. It is assumed that the vector potential $\vec{A}$ has the form $\vec{A} = (0, 0, A(x,t))$.

Let us use standard expression for the current density:

$$j = e\sum_{ps} v_s(p - \frac{e}{c}A(t) - \frac{e}{c}A_0(t))\langle a_{ps}^+ a_{ps}\rangle, \qquad (4)$$

where $v_s(p) = \dfrac{\partial \varepsilon_s(p)}{\partial p}$ and brackets mean averaging with the density nonequilibrium matrix $\rho(t)$: $\langle B \rangle = Sp(B(0)\rho(t))$. Then from the motion equation for the density matrix noting that $[a_{ps}^+ a_{ps}, H] = 0$ we get $\langle a_{ps}^+ a_{ps}\rangle = \langle a_{ps}^+ a_{ps}\rangle_0$, where $\langle B \rangle_0 = Sp(B(0)\rho(0))$. And then noting that $\rho_0 = \exp(-H/kT)/Sp(\exp(-H/kT))$ (k is the Boltzmann's constant, T is the temperature) and summing all the mentioned we obtain an exact equation for the electric field vector potential:

$$\frac{\partial^2 \vec{A}}{\partial x^2} - \frac{1}{c^2}\frac{\partial^2 \vec{A}}{\partial t^2} + \frac{8\pi e \gamma a}{c} \sin\left(\frac{eaA}{c} + \frac{eaA_0(t)}{c}\right) \bullet$$

$$\sum_{s=1}^{m} \int_{-\pi/a}^{\pi/a} dp \frac{\cos(ap)\cos(\pi s/m)}{\sqrt{1 + 4\cos(ap - aeA/c - eaE_o t)\cos(\pi s/m) + 4\cos^2(\pi s/m)}} \frac{\exp(-\beta \varepsilon_s(p))}{1 + \exp(-\beta \varepsilon_s(p))} = 0 \quad (5)$$

$$\beta = 1/kT$$

After the equation (5) is non-dimensionalized and square root is expanded into cosine Fourier series $\cos(ap - aeA/c - aeA_0(t)/c)$ and the integration is completed this equation can be represented by the following relationships:

$$\frac{\partial^2 B}{\partial x'^2} - \frac{1}{c^2}\frac{\partial^2 B}{\partial t'^2} + \sin(B + A_1(t)) + \sum_{k=2}^{\infty} b_k \sin(k(B + A_1(t))) = 0$$

$$B = \frac{eaA}{c}; x' = \frac{ea}{c}\sqrt{8\pi\gamma\delta}; t' = t\frac{ea}{c}\sqrt{8\pi n_0 \gamma \delta}; A_1(t) = \frac{eaA_0(t)}{c} \quad (6)$$

$$\delta = \sum_{s=1}^{m} \int_{-\pi/a}^{\pi/a} dp \frac{\cos(ap)\cos(\pi s/m)}{\sqrt{1 + 4\cos^2(\pi s/m)}} \frac{\exp(-\beta \varepsilon_s(p))}{1 + \exp(-\beta \varepsilon_s(p))}$$

here $n_0$ is the electrons equilibrium density in the carbon nanotubes.

It should be noted that if $A_0(t) = const$ the equation (6) is a generalization of the well known sine-Gordon equation in case when a generalized potential can be expanded into Fourier series. Obviously the generalized potential arising in this problem is defined as $U(B) = \cos(B) + \sum_{k=2}^{\infty} b_k \cos(kB)$. The calculations have demonstrated that the form of this potential determining the properties of nonlinear solutions depends on temperature weakly.

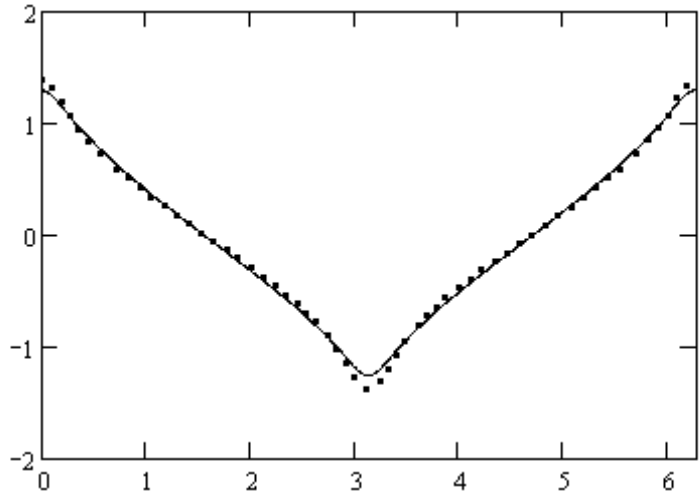

Fig.2. Dependence of the potential arising in the equation (6) on temperature. The value of B is plotted along the *x*-axis, the value of U(B) is plotted along the *y*-axis. The continuous curve indicates the temperature increased 20 times.

The potential under consideration depends on the type of carbon nanotubes more highly.

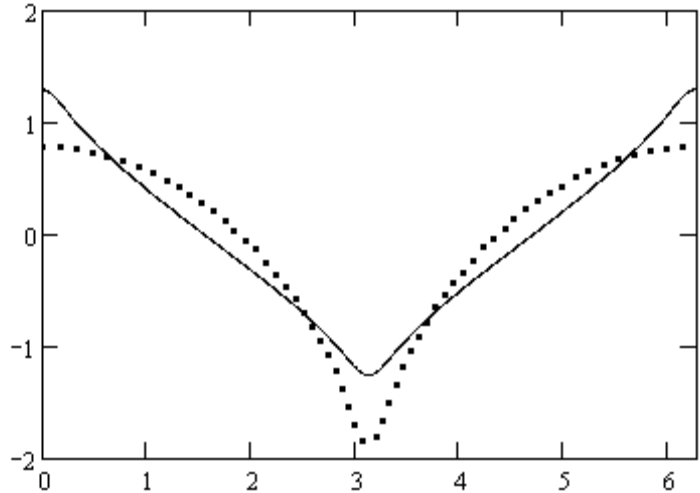

Fig.3. Dependence of the potential arising in the equation (6) on the type of carbon nanotubes. The value of B is plotted along the *x*-axis, the value of U(B) is plotted along the *y*-axis. The continuous curve indicates the (7, 0) nanotubes, the dotted curve indicates the (14, 0) nanotubes.

Due to coefficients $b_k$ decreasing with k increasing, we can retain the first two nonvanishing terms in the sum in the relationship (6). As a result we obtain the double sine-Gordon equation which is widely used in applications but cannot be integrated by the inverse scattering method [16]. An important consequence of this equation is the theorem of areas: only the pulses having a specific "area" are stable with respect to a change in the shape (the area of the pulse $\psi(t)$ is defined as $\int_{-\infty}^{\infty} \psi(t)dt$). The pulses having a larger area tend to decrease it to the specific value and, conversely, the pulses having a smaller area tend to increase it. Second [16], in case of rapidly decreasing boundary conditions the character of pulse interaction and what is the most important the character of the single pulse separation depend substantially on the pulse velocity. As the pulse velocity increases the pulses begin to interact more elastically and a smaller part of their energy transforms into vibrational modes.

These circumstances have given impetus to further numerical investigation of the equation (6), which has been obtained without any restrictions on the minimum duration of the electric field pulse.

**3. The results of numerical analysis. A case of null constant field.**

The equations under investigation were numerically solved according to the cross-type finite-difference scheme [17]. A constant field was assumed to be equal to zero on the first stage. The time and coordinate steps were determined from the standard conditions of stability. The steps of the finite-difference scheme were sequentially decreased by a factor of two until the solution changed in the eighth significant figure. An initial condition was chosen in the form of the well known kink-solution of the sine-Gordon equation:

$$B(x,t) = 4 arctg(\exp((x-vt)/\gamma))$$
$$\gamma = (1-v^2)^{1/2}$$

The given initial condition corresponds to the fact that the critical short pulse consisting of the single "half-oscillation" of the electric field is fed to the sample. The following values of parameters $\gamma \approx 2.7$ eV, $b = 0.142$ nm were chosen based on the problem geometry and also were estimated for instance by using the quantum-chemical semiempirical method MNDO [18] according to the CNT electronic structure calculations [19].

An arising evolution of the electromagnetic field in time is shown in Fig.4.

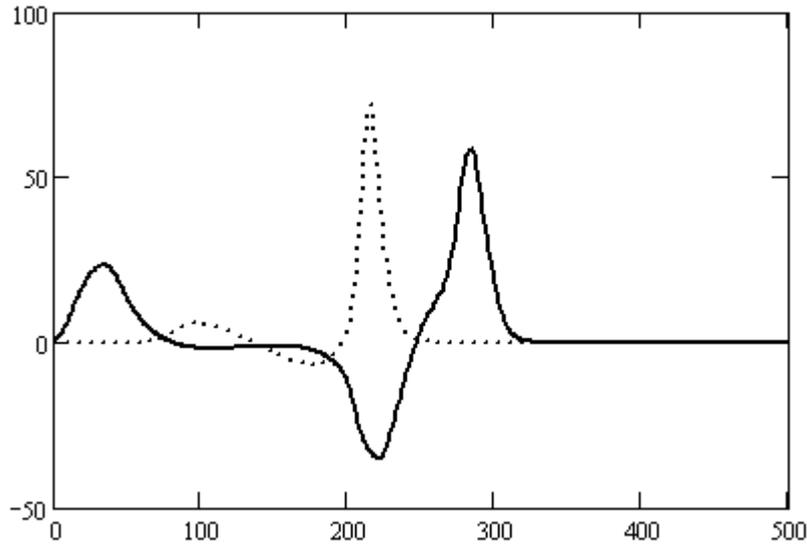

Fig.4. Dependence of the electric field determined by the potential in the equation (6) on coordinate at different instants of time. The non-dimensionalized coordinate (the unit is equal to $3 \bullet 10^{-8}$ m) is plotted along the *x*-axis, the non-dimensionalized value of alternating electric field (the unit is equal to $10^7$ V/m) is plotted along the *y*-axis. The continuous line indicates the time is 2 times greater than the dotted one. v/c=0,95.

It can be seen that the ultrashort pulse is decomposed into two pulses with the substantially different amplitudes. It should be noted that a similar behavior was observed in studying an analog of the sine-Gordon equation in other nonlinear systems [20]. It is worth noting that the ultrashort pulse "throws off" its extra area which is separated from the pulse. That in general is in good agreement with the investigation results of systems evolution described by an equation close to the sine-Gordon equation. As was noted above the former circumstance is associated with the fact that the system under consideration can be adequately described in terms of the double sine-Gordon equation for which there is an analog of the theorem of areas [20].

Typical dependences of a non-dimensionalized according to the equation (6) electric field amplitude on coordinate in case of the same initial conditions but corresponding to the different carbon nanotubes structures are illustrated in Fig.5.

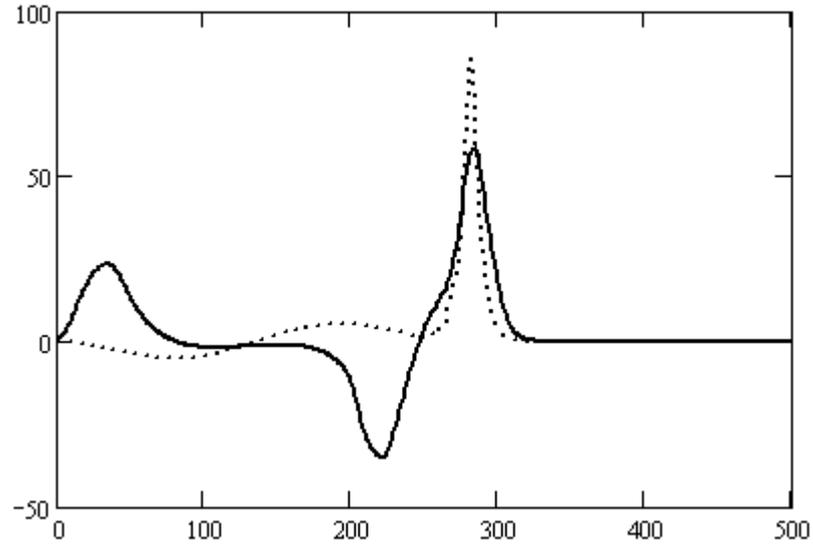

Fig.5. Dependence of the electric field determined by the potential in the equation (6) on coordinate at a fixed time moment. The non-dimensionalized coordinate (the unit is equal to $3 \bullet 10^{-8}$ m) is plotted along the *x*-axis, the non-dimensionalized value of the alternating electric field (the unit is equal to $10^7$ V/m) is plotted along the *y*-axis. The continuous curve indicates the (7, 0) nanotubes, the dotted curve indicates the (14, 0) nanotubes.

It should be noted that in nanotubes with great number of atoms along the circumference a widening of the electromagnetic-field pulse is smaller during its propagation. It can be directly attributed to the circumstance that the more nanotubes diameter the more effective nonlinearity is (see the Fig.3), which prevents the spreading dispersion pulse according to the equation (6). As a result the nanotubes with greater diameter are more perspective in devices using the ultrashort optical pulses. Besides such behavior can be explained by a simple mechanical analogy. If it is taken into consideration that equation (6) in static case is equivalent to the problem of the mass motion in an electric field with the potential defined by

$$\frac{\partial U(B)}{\partial E} = \sin(B) + \sum_{k=2}^{\infty} b_k \sin(kB)$$

then since the potential well for the nanotubes of major diameter is deeper the particle will be in the area limited by the well sides the basic part of time. That is why the electromagnetic pulse will be shorter.

**4. The results of numerical analysis. A case of external electric fields.**

In case of nonzero constant field calculations were completed as in the previous article by using the same values. An arising evolution of alternating electro-magnetic field in time is shown in Fig.6.

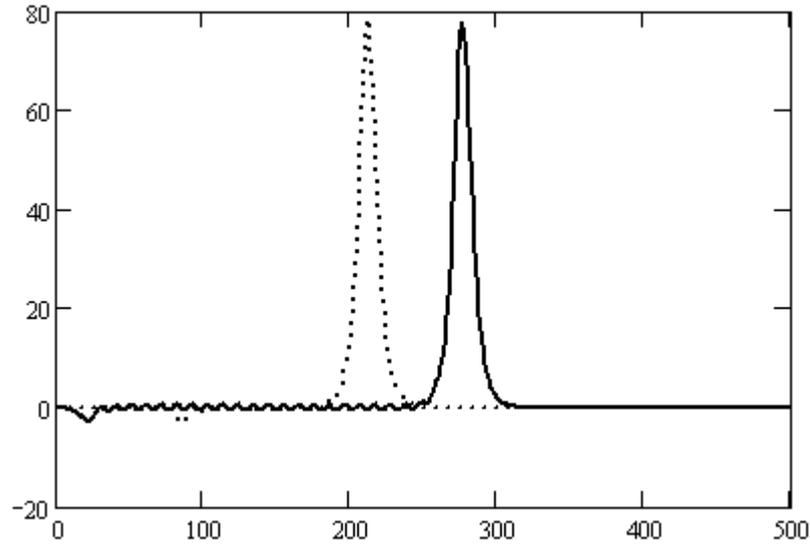

Fig.6. Dependence of the electric field determined by the potential in the equation (6) on coordinate at fixed time moment. The non-dimensionalized coordinate (the unit is equal to $3 \bullet 10^{-8}$ m) is plotted along the *x*-axis, the non-dimensionalized value of the alternating electric field (the unit is equal to $10^7$ V/m) is plotted along the *y*-axis. The constant field is $10^6$ V/m. The continuous line indicates the time is 2 times greater than the dotted one. v/c=0,95.

It should be noted that the external constant field exerts antihunting action and shrinks the alternating electromagnetic-field pulse in comparison to the constant field-free case. It can be associated with the fact that in the presence of the constant field in an electron spectrum so called "Stark ladder" is arisen and electrons can vary their energy only by the value proportional to the difference between adjacent energy levels of the "Stark ladder". This leads to a decrease in an effective electrons dispersion which means in its turn an increase in a dispersion spreading of the alternating electric field pulse.

The calculations have demonstrated that in case of the applied constant electric field the pulse shrinking depends on the type of the carbon nanotubes weakly and is mainly determined by the value of the external constant electric field. It should be noted that the alternating electric-field pulse starting with some value of the electric field ceases to shrink and begins to propagate in a stationary state. Such behavior can be explained qualitatively both by a competition between processes of the pulse dispersion propagation in a field-free case and electrons motion along the "Stark ladder".

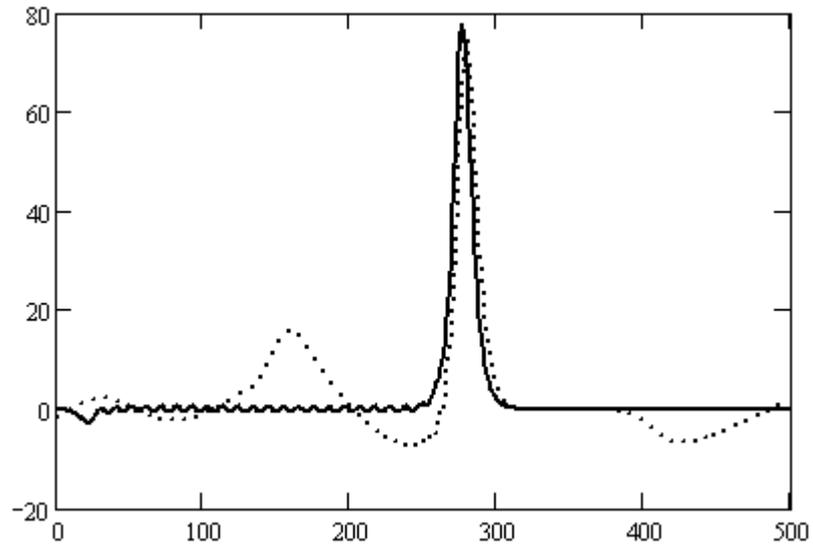

Fig.7. Dependence of the electric field determined by the potential in the equation (6) on coordinate at a fixed time moment. The non-dimensionalized coordinate (the unit is equal to $3 \bullet 10^{-8}$ m) is plotted along the *x*-axis, the non-dimensionalized value of the alternating electric field (the unit is equal to $10^7$ V/m) is plotted along the *y*-axis. The continuous line indicates the constant field of the value $10^6$ V/m, and $10^3$ V/m is for dotted line. v/c=0,95.

The very competition between these processes leads to the alternating electric field dynamics demonstrated in Fig.7. Notice that oscillations preceding to pulse front are attributed to the electrons system response to the constant field which leads to periodic oscillations determined by the periodic character of the electrons dispersion law (2).

Thus, the applied external constant electric field decreases the duration of the critical short pulse of the alternating electric field and prevents its decomposing into smaller pulses.

In case of the applied external harmonic electric field the main qualitative dependences are preserved (see Fig.8). And as in the case of the constant field there is the pulse shrinking. Notice that in case of the alternating field there is an increase in pulse amplitude.

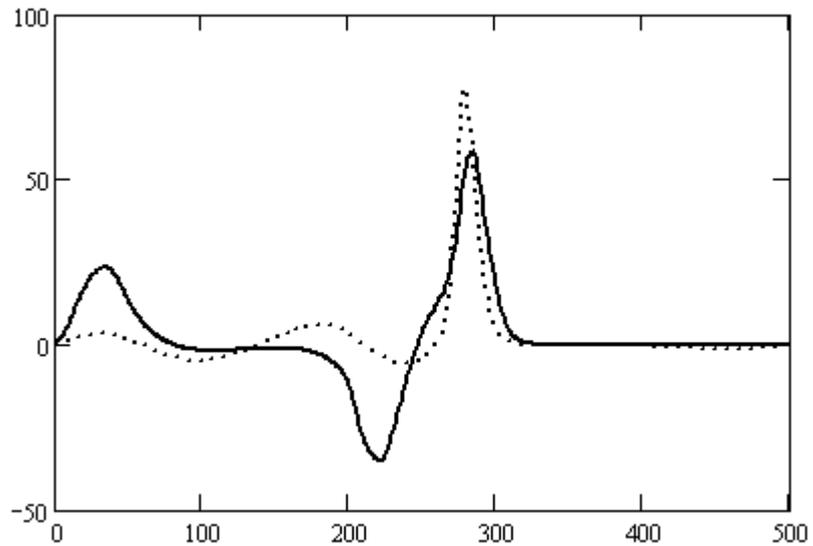

Fig.8. Dependence of the electric field determined by the potential in the equation (6) on coordinate at a fixed time moment. The non-dimensionalized coordinate (the unit is equal to $3 \cdot 10^{-8}$ m) is plotted along *x*-axis, the non-dimensionalized value of the alternating electric field (the unit is equal to $10^7$ V/m) is plotted along the *y*-axis. The continuous curve indicates the absence of the external field, the dotted curve indicates the external harmonic electric field with amplitude $10^5$ V/m, and frequency $10^{12}$ Hz.

In our opinion such behavior can be associated with the fact that the external alternating electric field prevents the ultrashort pulse decomposition into a number of smaller pulses. The results of numerical calculations have demonstrated this phenomena depends on the alternating field frequency mostly (see Fig.9).

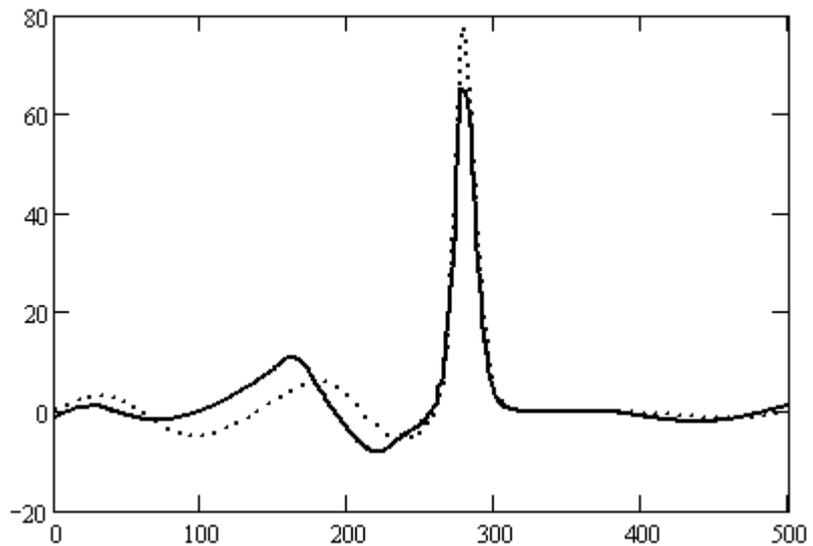

Fig.9. Dependence of the electric field determined by the potential in the equation (6) on coordinate at a fixed time moment. The non-dimensionalized coordinate (the unit is equal to $3 \cdot 10^{-8}$ m) is plotted along *x*-axis, the non-dimensionalized value of the alternating electric field (the unit is equal to $10^7$ V/m) is along the *y*-axis. The external electric field has amplitude $10^5$ V/m. The continuous curve indicates the external field frequency $10^{10}$ Hz, and $10^{12}$ Hz is for dotted one.

In conclusion we present data on the phase modulation injection into the external constant field. If the external field is given by

$$E_0(t) = E_0 \cos(wt + at^2)$$

then the modulation injection provided by summand proportional to *a* leads to additional increase in the pulse amplitude (see Fig.10).

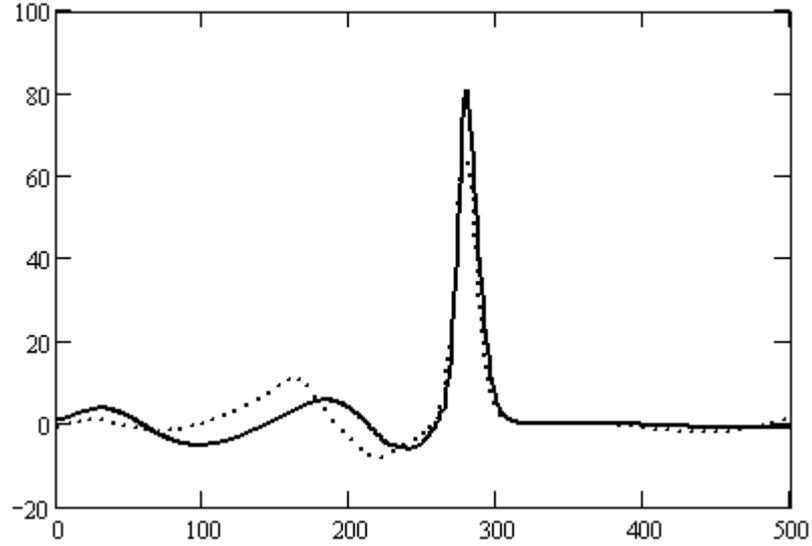

Fig.10. Dependence of the electric field determined by the potential in the equation (6) on coordinate at a fixed time moment. The non-dimensionalized coordinate (the unit is equal to $3 \bullet 10^{-8}$ m) is plotted along *x*-axis, the non-dimensionalized value of the alternating electric field (the unit is equal to $10^7$ V/m) is along the *y*-axis. The continuous curve indicates the phase modulation injection, the dotted curve indicates modulation free. Parameters are the same as in the Fig.9.

The above specific feature can be explained similarly by the laser's active mode locking phenomena. The external alternating field establishes the phase relations between the carbon nanotubes electron modes which differ by *s* in the quasi-momentum $\vec{p} = (p, s)$. Hence the phase modulation injection leads to more effective relation, which results in a pulse amplitude increase.

**5. Conclusion**

The results of the performed investigation allow us to make the following main conclusions:

1. The equation describing the dynamics of electromagnetic field in a carbon nanotubes system in case of low temperatures and for the critical short pulses was obtained.

2.  The solutions of the efficient equation which are the analog of the sine-Gordon equation solitons were obtained.
3.  In general case an initial disturbance falls into several pulses, and some of them move in the direction opposite to the pulse motion.
4.  As a result of evolution the pulses of the carbon nanotubes of greater radius attain less duration than the carbon nanotubes of a smaller radius (at the equal initial duration). It can be associated with the major effective nonlinearity of the carbon nanotubes of a greater radius.
5.  An applied constant electric field prevents an increasing in the electromagnetic pulse duration and "suppresses" its widening on account of dispersion.
6.  The electromagnetic pulse propagation in bundles of carbon nanotubes in a presence of constant electric field has a stable character and the carbon nanotubes of a greater radius can be used in producing very short pulse devices.
7.  An external alternating electric field does not only prevent an increase in the alternating electric-field pulse duration and "suppresse" its widening in consequence of dispersion but also leads to increase in the alternating field pulse amplitude.
8.  The phase modulation injection into the alternating external electric field leads to additional increase in the alternating electric-field pulse amplitude.